\documentclass[prb,aps,twocolumn,amstex,preprintnumbers,amsmath,amssymb,array,ta
bularx]{revtex4}

\usepackage{graphicx}
\usepackage{dcolumn}
\usepackage{bm}

\DeclareGraphicsExtensions{.jpg,.pdf,.eps}

\begin{document}
\title{Magnetoresistance Devices Based on Single Walled Carbon
Nanotubes}

\author{$\mbox{Oded Hod}^{1}$, $\mbox{Eran Rabani}^{1}$ and $\mbox{Roi
Baer}^{2}$}

\address{$^{1}$School of Chemistry, Tel Aviv University, Tel Aviv.
69978, Israel\\ $^{2}$Institute of Chemistry and Lise Meitner Center
for Quantum Chemistry, the Hebrew University of Jerusalem, Jerusalem
91904 Israel}

\date{\today}

\begin{abstract}
We demonstrate the physical principles for the construction of a
nanometer sized magnetoresistance device based on the Aharonov-Bohm
effect.  The proposed device is made of a short single-walled carbon
nanotube (SWCNT) placed on a substrate and coupled to a tip.  We
consider conductance due to motion of electrons along the
circumference of the tube (as opposed to motion parallel to its axis).
We find that the circumference conductance is sensitive to magnetic
fields threading the SWCNT due to the Aharonov-Bohm effect, and show
that by retracting the tip, so that its coupling to the SWCNT is
reduced, very high sensitivity to the threading magnetic field
develops. This is due to the formation of a narrow resonance through
which the tunneling current flows.  Using a bias potential the
resonance can be shifted to low magnetic fields, allowing the control
of conductance with magnetic fields of the order of $1$~Tesla.  PACS:
73.63.-b, 73.63.Fg, 75.75.+a
\end{abstract}

\maketitle
 
\newpage 
Understanding nanoscale electronic devices is intertwined with the
ability to control their properties. One of the most scientifically
intriguing and potentially useful property is the control of the
electrical conductance in such devices.\cite{Nitzan03,Joachim00} One
convenient way of affecting conductance is by applying magnetic
fields. In mesoscopic systems, for example, the conductance is
sensitive to the Aharonov-Bohm (AB) effect.\cite{Aharonov59} The study
of the interplay between bias and gate voltages, and magnetic fields
in these systems has lead to development of micronic AB
interferometers.\cite{Webb85,Yacoby95,Shea00} At the nanoscale,
however, it is widely accepted that AB interferometers do not
exist.\cite{Hod04} This is because unrealistic huge magnetic fields
are required to affect conductance through a loop encircling very
small areas (for a loop of area $A$, the magnetic field needed to
complete a full AB period can be obtained from the relation $A B =
\phi_{0}$, where $\phi_{0}=h/e$ is the flux quantum, $h$ and $e$ are
Planck's constant and electron charge, respectively).  Thus, at the
nanoscale, devices that exhibit large magnetoresistance have been
demonstrated based on the Zeeman spin splitting of individual
molecular states,\cite{McEuen02} or based on the Kondo
effect.\cite{Park02}

Recently, the AB effect has been measured for single-walled and
multi-walled carbon nanotubes (SWCNT and MWCNT,
respectively).\cite{Bachtold99,Bezryadin04,Smally04} The MWCNT with
relatively large diameter ($15$nm), exhibits $h/e$-period magnetic
flux dependence with $B=5.8$~Tesla for a full AB period.  This result
is important, showing that transport is coherent through the tube, in
agreement with previous observation.\cite{Park01} The measurements for
the smaller diameter SWCNT indicate that the band structure of the
tube depends on the magnetic flux threading it. But, a full AB period
and, thus, switching capability, would require much higher magnetic
fields than those used ($B_{max}=45$~Tesla).  Therefore, an open
problem is whether SWCNTs, which have been proposed as ideal
candidates for the fabrication of nanoelectronic
devices,\cite{Dai96,Smalley97,Zettl97,Srivastava98,Fuhrer00,Dekker01,Avouris02}
can exhibit large magnetoresistance (at small magnetic fields) despite
the fact that a magnetic field exceeding $1000$~Tesla is required to
complete a full AB period for a tube with a diameter comparable to
$1nm$.

\begin{figure}
\begin{center}
\includegraphics[width=6.5cm]{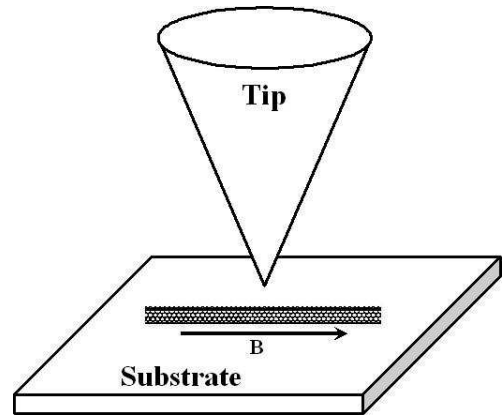}
\end{center}
\caption{An illustration of the experimental setup suggested for
measuring the cross sectional magnetoresistance of a CNT.}
\label{fig:illustration}
\end{figure}

In this letter, we layout the simple physical principles required to
overcome this problem.  Utilizing the AB effect, we suggest a way to
switch the conductance through the nanometric cross section of a SWCNT
by the application of small ($\approx 1$~Tesla) magnetic fields
parallel to the axis of the tube.  Our scheme also provides a
framework to study and control coherent transport in SWCNTs.

We considered a SWCNT placed on a conducting substrate coupled to a
Scanning Tunneling Microscope (STM) tip from above as described
schematically in Fig.~\ref{fig:illustration}.  A bias potential is
applied between the STM tip and the underlying surface.  We calculate
the current through the circumference of the SWCNT (and not along the
tube axis).  When a magnetic field is applied perpendicular to the
cross section of the tube (along its main axis), electron pathways
transversing the circular circumference in a clockwise and a
counterclockwise manner gain different magnetic phases, and thus AB
interference occurs.

To calculate the conduction between the STM tip and the substrate in
the presence of a magnetic field applied, we have developed a simple
approach based on the Magnetic Extended Huckle Theory
(MEHT).\cite{Hod04} Within this approach, we add the proper magnetic
terms to the Extended Huckle (tight binding) Hamiltonian,
$\hat{H}_{EH}$ (from now on we use atomic units, unless other wise
noted):
\begin{equation}
\hat{H} = \hat{H}_{EH} - \mu _B \hat{\bf{L}}\cdot {\bf B} +
\frac{B^2}{8} R_\perp^2.
\label{eq:Hamiltonian}
\end{equation}
Here $\mu _B=\frac{1}{2}$ is the Bohr magneton in atomic units,
$\hat{\bf{L}}$ is the angular momentum operator, ${\bf B}$ is the
magnetic field vector, and ${\bf R}_\perp$ the projection of ${\bf R}$
onto the plane perpendicular to ${\bf B}$. A gauge invariant Slater
type orbitals (GISTO) basis set is used to evaluate the MEHT
Hamiltonian matrix:
\begin{equation}
|GISTO\rangle _{\alpha} = |STO\rangle _{\alpha}
 e^{-\frac{ie}{\hbar}{\bf A}_{\alpha}\cdot {\bf r}},
\label{eq:GISTO}
\end{equation}
where $|STO\rangle _{\alpha}$ is a Slater type orbital centered on
atom ${\alpha}$, and ${\bf A}_{\alpha}$ is the vector potential
evaluated at the nuclear position ${\bf R}_{\alpha}$, ${\bf
A}_{\alpha} = -\frac{1}{2}({\bf R}_{\alpha} \times {\bf B}_{\alpha})$.

For the results reported below, all carbon atoms of the tube were
treated explicitly.  In the MEHT, each carbon atom contributes two
$s$-electrons and two $p$-electrons.  The valence $s$ and $p$ atomic
orbitals are explicitly considered in the Hamiltonian.  The STM tip
and the contact between the tube and the substrate are modeled by a
one dimensional atomistic conducting wire (see more details below).
We assume a homogeneous magnetic field parallel to the tube axis and
calculate the Hamiltonian matrix elements analytically~\cite{Hod04a}
within the Pople approximation.\cite{Pople62}

The conductance is calculated using the Landauer
formalism~\cite{Landauer57} which relates the conductance to the
scattering transmittance probability through the system:
\begin{equation}
g=g_0\frac{\partial}{\partial V}\int
[f_T^{\frac{1}{2}V}-f_S^{-\frac{1}{2}V}]T(E)dE.
\label{eq:Landauer}
\end{equation}
In the above equation $g_0=2e^2/h$ is the quantum conductance,
$f_{T/S}^{\pm \frac{1}{2}V}(E)=[1+e^{\beta (E-\mu _{T/S}\pm
\frac{1}{2}V)}]^{-1}$ is the Fermi-Dirac distribution in the STM
tip/substrate, $\beta =(k_{\mbox{\tiny {B}}} T)^{-1}$ is the inverse
temperature, $\mu_{T/S}$ is the chemical potential in the STM
tip/substrate, and $V$ is the bias potential.  We assume that the bias
potential across the system drops sharply at both
contacts.\cite{Nitzan03a} The transmittance $T(E)$ is given
by~\cite{Baer03}
\begin{equation}  
T=4tr\{ \hat{G}^\dagger \Gamma_T \hat{G} \Gamma_S\}.
\label{eq:t}
\end{equation}
Here, $\Gamma _{T/S}$ is the imaginary absorbing potential
representing the imaginary part of the self-energy ($\Sigma$) of the
STM tip/substrate (we assume that the real part of $\Sigma$ is zero),
and $\hat{G}(E)=[E-\hat{H}+i(\Gamma_T + \Gamma_S)]^{-1}$ is the
appropriate Green function.  In the following calculations, the
imaginary potential was taken in the form of a Gaussian centered at
$b_{T/S}$: $\Gamma _{T/S}=iV_0\exp\{-\frac{(z-b_{T/S})^2}{2\sigma
^2}\}$ where the height of the potential $V_0\approx
\epsilon_F-\epsilon_0$ was approximated by the height of the Fermi
energy ($\epsilon_F$) above the valence band bottom ($\epsilon_{0}$).
The width of the absorbing potential was $\sigma \approx 10$\AA.

\begin{figure}
\begin{center}
\includegraphics[width=8cm]{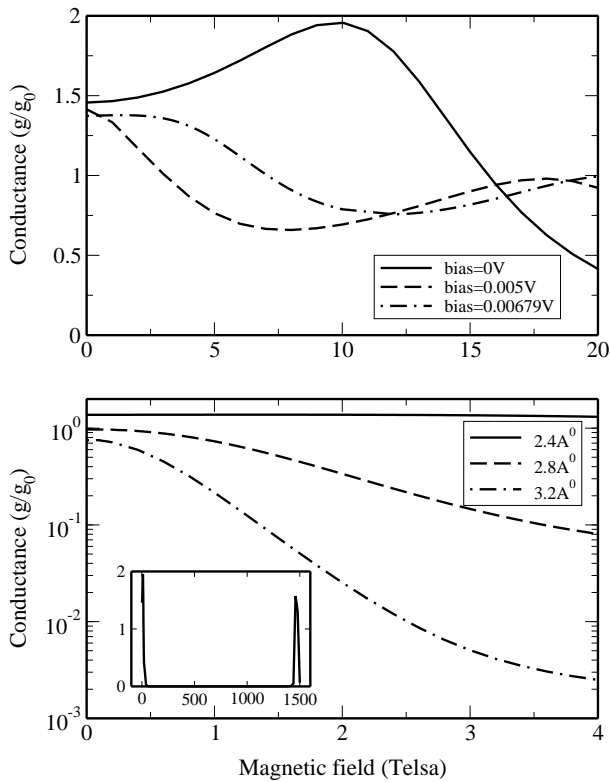}
\end{center}
\caption{Conductance versus the magnetic field for a $24\mbox{x}0$
SWCNT. Upper panel: The effect of bias potential on the position of
the conductance peaks at tube-tip separation of $2.4$\AA. Lower panel:
The effect of increase in the tube-tip separation at a constant bias
potential of $0.00679V$. Inset: The full AB period for a $24\mbox{x}0$
SWCNT at zero bias potential and tube-tip separation of $2.4$\AA.}
\label{fig:G_B_24x0x4}
\end{figure}

In Fig.~\ref{fig:G_B_24x0x4}, the conductance through the cross
section of a $24\mbox{x}0$ SWCNT is plotted against the external axial
magnetic field applied.  These calculations were done for a tube four
unit cells in length.  Tests on longer tubes reveal the same
qualitative picture described below.  We use minimum image periodic
boundary conditions for the passivation of the edge atoms.  At the
inset of the lower panel of Fig.~\ref{fig:G_B_24x0x4}, the full AB
conductance period is plotted under zero bias voltage and a separation
of $2.4$\AA~between the STM tip/substrate and the tube.  As can be
seen, the full period equals $\sim 1500$~Tesla, which is expected for
an AB interferometer with a radius comparable to $1$~nm.

In the upper panel of Fig.~\ref{fig:G_B_24x0x4}, we plots the
circumference conduction as a function of the axial magnetic field at
low values of the field, and for several values of the bias potential.
The circumference conduction at zero bias first increases as we switch
on the magnetic field (negative magnetoresistance), peaks near
$B=10$~Tesla, and decreases to zero above $B=30$~Tesla.  The maximum
conduction observed $g/g_{0}=2$ is limited by our conducting wires.
In order to achieve switching capability at magnetic fields smaller
than $1$~Tesla, it is necessary to move the conduction peak to zero
magnetic fields and at the same time reduce the width of this peak.

When a small bias is applied to the sample the conduction peak splits
into a doublet.  This split can be associated with the specific choice
of the potential drop across our system.  The position of the
corresponding peaks depends on the value of the bias.  As can be seen
in the figure, by adjusting the bias potential it is possible to shift
one of the conductance peaks toward low values of the magnetic field,
such that the conductance is maximal at $B=0$ and positive
magnetoresistance is achieved.  The shift in the conductance peak can
be attributed to the change in the energy level through which
conductance occurs when a small bias is applied.  As a results of this
change, the electron momentum changes and this leads to a shift seen
in the conduction peak.  We return to this point below when we analyze
the results in terms of a simple continuum model.

In the lower panel of Fig.~\ref{fig:G_B_24x0x4} the effect of changing
the tube-tip/substrate separation at constant bias potential is
studied.  As one increases the separation between the tube and the
tip/substrate, the coupling between the tube and the tip/substrate
decreases, resulting in a reduction of the width of the energy
resonances of the SWCNT.  Thus, conduction becomes very sensitive to
an applied magnetic field and small variations in the field shift the
relevant energy level out of resonance.  In the magnetoresistance
spectrum, this is translated to a narrowing of the transmittance peaks
as can be seen in the lower panel of Fig.~\ref{fig:G_B_24x0x4}.  At
the highest separation studied ($3.6$\AA), the width of the
conductance peak is comparable to $1$~Tesla.

Considering the combined effect of the bias potential and the
tube-tip/substrate separation, it is possible to shift the position of
the conduction peak to small magnetic fields while at the same time
reducing its width.  This is achieved by carefully selecting the
values of the bias potential and tube-tip/substrate separation.  Under
proper conditions, we obtain positive magnetoresistance with a sharp
response occurring at magnetic fields comparable to $1$~Tesla.  This
result is significant since it implies that despite the fact that the
tube radius is small ($\sim 1$~nm) and the corresponding full AB
period requires unrealistic large magnetic fields ($\sim 1500$~Tesla),
it is possible to achieve magnetic switching at relatively small
magnetic fields.  This result also agrees with recent experiments that
show that the band structure of a carbon nanotube depends on the
magnetic flux threading it.\cite{Smally04}

\begin{figure}
\begin{center}
\includegraphics[width=8cm]{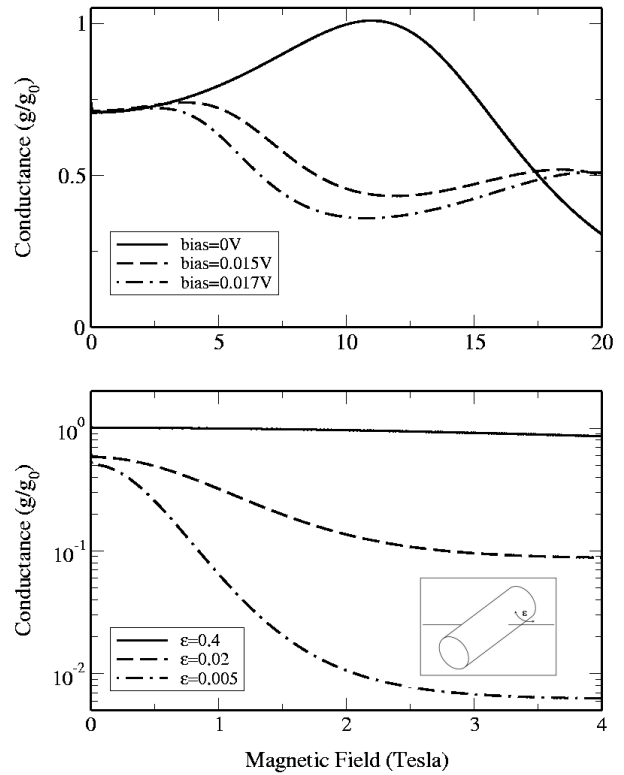}
\end{center}
\caption{Conductance versus the magnetic field as computed from the
continuum model. The wave length of the Fermi electrons is taken to be
$\lambda _F=2\pi /k_F=3.522$\AA and the diameter of the tube is the
same as that of the $24\mbox{x}0$ SWCNT ($\sim 2$~nm). Upper panel:
For a given value of the junction scattering amplitude $\epsilon
=0.0025$ the application of a bias potential splits the conductance
peak and shifts the peaks along the magnetic field axis. Lower panel:
At a constant bias voltage ($0.0215$~V) reducing the scattering
parameter results in a narrowing of the transmittance peak. Inset: an
illustration of the 1D model setup. The tip and substrate are
represented by conducting wires.}
\label{fig:G_B_analytic}
\end{figure}

We now turn to analyze the results described above in terms of a
simple continuum model.  We show that they are reproducible with a
single adjustable parameter.  Our model includes a cylindrical tube of
cross section $A$ coupled to two leads as sketched in the inset of
Fig.~\ref{fig:G_B_analytic}. The leads mimic the effect of the tip and
substrate.  A magnetic field of strength $B$ is applied along the tube
axis, and the threading magnetic flux is therefore $\phi = A B$.

The transmission probability $T(\phi)$ that a conducting electron with
energy $E_{k}$ originating in the left wire (the STM tip) will pass
through the loop and emerge at the right wire (the substrate) can be
calculated exactly.\cite{Gefen84,Hod04} Assuming coherent scattering
at the incoming and outgoing junctions, we obtain:
\begin{equation}
T(\phi) = T_0 \frac{1+\cos (2\pi\phi/\phi_{0})} {1+P/R\cos
(2\pi\phi/\phi_{0}) + Q/R \cos (4\pi\phi/\phi_{0})}.
\label{eq:transmittance}
\end{equation}
In the above equation, $T_0=32R\epsilon ^4\sin^2(\theta _k)$, as
before $\phi_{0}=\frac{h}{e}$ is the quantum flux, and $P, R, Q$ are
given by $P=2(c+1)^2[(c-1)^2-2(c^2+1)\cos 2\theta _k]$,
$R=(c-1)^4+4[c^4+2c^2\cos 4\theta _k+1-(c^2+1)(c-1)^2\cos 2\theta _k]$
and $Q=(c+1)^4$ where $c=\sqrt{1-2|\epsilon |^2}$.  These coefficients
depend on two independent parameters: the amplitude $\epsilon$ for an
electron to scatter into the tube from the tip, and the spatial phase
accumulated by an electron transversing half the circumference of the
tube $\theta_{k}= k L / 2$.  Here, $k=\sqrt{2m^{*} E_{k}/\hbar ^2}$ is
the wave number of the conducting electron, $L$ is the circumference
of the tube, and $m^{*}$ is the effective mass of the electron.
Conduction is obtained by inserting the result for the transmittance
(cf. Eq.~(\ref{eq:transmittance})) into the Landauer formula given by
Eq.~(\ref{eq:Landauer}).

In Fig.~\ref{fig:G_B_analytic}, we plot the conduction calculated for
the continuum model. The value of $\theta_{k}=kL/2$ was calculated
with the wave number approximately equal to that of a Fermi electron
in a graphene sheet ($k=2\pi/3.52 \AA^{-1}$), and $\epsilon$ was
adjusted to reproduce the results for the SWCNT shown in
Fig.~\ref{fig:G_B_24x0x4}.  Analyzing the expression for the
transmittance through the tube (cf. Eq.~(\ref{eq:transmittance})), we
find that the position of the conducting peaks depends mainly on the
value of ($\theta_{k}$). For a maximum conduction at small magnetic
fields, $\theta_{k}=n\pi$, where $n=1,2,3 \cdots$.  Thus for a given
tube circumference $L$, the wave number of the conducting electrons
must satisfy $k=2\pi n/L$.  This is not the case for the parameters
chosen above and the maximum conduction occurs at $B=10$~Tesla.  Thus,
in order to move the conduction peak to $B=0$ we apply a bias
potential.  Introducing a bias potential changes the energy level
through which conductance occurs, resulting in a change in the wave
number ($k=\sqrt{2 m^{*}E_{k}/\hbar^2}$) and the spatial angle
($\theta_{k}=kL/2$).  In the upper panel of
Fig.~\ref{fig:G_B_analytic}, we plot the effect of adding a bias
potential on the conduction calculated within the continuum model.
Similar to the case for the SWCNT shown in the upper panel of
Fig.~\ref{fig:G_B_24x0x4}, the position of the peaks is sensitive to
the value of the bias potential.

In the lower panel of Fig.~\ref{fig:G_B_analytic}, we plot the
conductance for different values of the junction scattering parameter
$\epsilon$ (the only free parameter in our theory).  The value of the
bias potential is chosen such that conduction peaks at zero magnetic
field, and positive magnetoresistance is achieved.  As can be seen in
the figure, the width of the conductance peaks decreases upon a
decrease in $\epsilon$.  This effect is identical to that observed for
SWCNTs when the tube-tip/substrate separation increased. The decrease
in $\epsilon$ results in a decrease of the width of the energy levels
on the tube, since the coupling to the continuum leads is reduced.
Thus, for low values of $\epsilon$ small magnetic fields are
sufficient to move the conducting level out of resonance, and
conduction becomes very sensitive to the value of the magnetic field.

The above calculations assume a low temperature of $1$~K. However, the
effects we report will hold even at higher temperatures.  The
temperature $T$ must be low enough to resolve the magnetic field
splitting of circumference energy levels, and must satisfy $k_{B}T <
\left[\hbar^{2} k_{f}/ m^{*} D\right] \left[\phi /
\phi_{0}\right]$. Here $D$ is the diameter of the tube, $k_{f}$ is the
Fermi wave number, and $m^{*}$ is the electron's effective mass.  For
a ratio of $\left[\phi / \phi_{0}\right] = 1/1500$ (namely, switching
capability at $1$~Tesla) the upper limit for the temperature is $\sim
20$~K.

In summary, we have demonstrated that SWCNT can be used as
magnetoresistance switching devices based on the AB effect.  The
essential procedure is to weakly couple the SWCNT to the conducting
tip/substrate in order to narrow the conducting resonances, while at
the same time to control the position of the resonances by the
application of a bias potential.  The fact that the diameter of the
tube is small becomes beneficial, since the separation between the
circumferential energy levels on the tube is large, and conductance
can be achieved through a single state.

This work was supported by The Israel Science Foundation (ER) and by
the German Israeli Science Foundation (RB).


\end{document}